# Observation of photonic antichiral edge states

Peiheng Zhou,[1] Gui-Geng Liu,[2,3,#] Yihao Yang,[2,3] Yuan-Hang Hu,[1] Sulin Ma,[1] Haoran Xue,[2] Qiang Wang,[2] Longjiang Deng,[1,*] and Baile Zhang[2,3,‡]

[1]*National Engineering Research Center of Electromagnetic Radiation Control Materials, State Key Laboratory of Electronic Thin Film and Integrated Devices, University of Electronic Science and Technology of China, Chengdu 610054, China*

[2]*Division of Physics and Applied Physics, School of Physical and Mathematical Sciences, Nanyang Technological University, 21 Nanyang Link, Singapore 637371, Singapore*

[3]*Centre for Disruptive Photonic Technologies, The Photonics Institute, Nanyang Technological University, 50 Nanyang Avenue, Singapore 639798, Singapore*

Chiral edge states are a hallmark feature of two-dimensional topological materials. Such states must propagate along the edges of the bulk either clockwise or counterclockwise, and thus produce oppositely propagating edge states along the two parallel edges of a strip sample. However, recent theories have predicted a counterintuitive picture, where the two edge states at the two parallel strip edges can propagate in the same direction; these anomalous topological edge states are named as antichiral edge states. Here we report the experimental observation of antichiral edge states in a gyromagnetic photonic crystal. The crystal consists of gyromagnetic cylinders in a honeycomb lattice, with the two triangular sublattices magnetically biased in opposite directions. With microwave measurement, unique properties of antichiral edge states have been observed directly, which include the titled dispersion, the chiral-like robust propagation in samples with certain shapes, and the scattering into backward bulk states at certain terminations. These results extend and supplement the current understanding of chiral edge states.

Topological materials in two dimensions (2D) are known to exhibit unidirectional, or chiral, edge states at the edges of the bulk [1-7]. For example, in the celebrated Haldane model that exhibits the quantum Hall effect with broken time-reversal symmetry [3], chiral edge states must propagate along the edges either clockwise or counterclockwise, depending on the sign of Chern number. As a consequence, a typical strip sample must support counter-propagating edge states along its two parallel edges. The time-reversal-invariant quantum spin Hall effect can be treated as a superposition of two copies of Haldane model, in which the chiral edge states emerge when the spin is fixed [4, 5]. Such chiral edge states have not only revolutionized condensed matter physics [1-5], but also opened a new chapter in photonics [6, 7] and acoustics [8], with applications ranging from one-way waveguides [9-12], robust optical delay lines [13], topological lasers [14-17], and photonic circuits [18, 19]. For all the above systems, the chiral edge states in a given strip geometry must propagate in opposite directions at the two parallel edges, and the numbers of left- and right-moving edge states must be equal.

Recent theories [20] have predicted another possibility: the edge states at the two parallel edges can propagate in the same direction, as shown in Fig. 1(a), being fully distinctive to conventional chiral edge states, and thus are named as antichiral edge states. These antichiral edge states can only occur in gapless systems since they need the associated bulk states to propagate in the opposite direction as required by energy conservation. A series of platforms have been theoretically proposed that can potentially realize antichiral edge states, including graphene structures [21, 22], exciton polariton strips [23], Heisenberg ferromagnets [24], and gyromagnetic photonic crystals [25]. However, due to the challenging configuration with broken time-reversal symmetry, antichiral edge states have not been observed in reality.

Here we report the experimental observation of antichiral edge states in a microwave-scale gyromagnetic photonic crystal. Such photonic crystals incorporate gyromagnetic materials to break time-reversal symmetry, and have been utilized to demonstrate the photonic chiral edge states, giving rise to the emerging field of topological photonics [6, 7]. We further incorporate on-site modulation of magnetization in such a platform, and successfully observe unique properties of antichiral edge states that are still difficult in other platforms.

The challenge to realize antichiral edge states stems from their underlying mechanism that can be described by a modified Haldane model [20] [see Fig. 1(b)]. In the original Haldane model [3], the next-nearest-neighbour coupling picks a nonzero phase that accounts for the magnetic flux at the center of a unit cell. However, in the modified Haldane model, the next-nearest-neighbour coupling for different sublattices have opposite signs in phase, meaning that the two sublattices of A and B sites shall "feel" opposite magnetic fluxes.

In photonics, the Haldane model can be realized in a gyromagnetic photonic crystal under uniform external magnetic fields [9-12]. In other words, all gyromagnetic rods are uniformly biased by external magnetic fields. However, for the modified Haldane model, what is necessary is the precise on-side modulation of magnetization. We adopt a honeycomb lattice of gyromagnetic rods as schematically illustrated in Fig. 1(c). The rods are yttrium iron garnet (YIG) ferrite materials, similar to previous demonstrations based on the Haldane model [10-12]. However, in the current configuration, the rods in the sublattices of A and B sites need to be magnetically biased in opposite directions. Note that a similar lattice configuration has been reported recently in Ref. [25], which shows the feasibility of constructing photonic antichiral edge states.

Figure 1(d) shows the schematic of the realization of on-site magnetization modulation. The photo of a concrete sample is shown in Fig. 1(e). The crystal of gyromagnetic rods is placed in an air-loaded waveguide composed of two parallel copper plates. On-site magnetization modulation is then enforced by placing samarium cobalt (SmCo) permanent magnets that sandwich the waveguide with careful designs. The magnets need to be placed exactly above and below the gyromagnetic rods. Each gyromagnetic rod is sandwiched between a pair of magnets that have the same biasing direction to form a uniform magnetic field in-between in the $z$ direction. To fix the positions, the magnets are embedded in two additional copper plates at the top and bottom of the waveguide. By flipping the biasing direction of magnets at neighbouring sites, the opposite magnetic fluxes are uniformly applied on the sublattices A and B in the $x$-$y$ plane, with the same strength of 0.043 Tesla (see Supplemental Material [26] for the measured magnetic flux distribution).

Note that a square array of small holes are drilled through the top copper plates to allow the insertion of probes for measurement [27]. The thickness of waveguide plates is essential (1 mm in our case), since it helps to tune the uniform distribution of magnetic fluxes. The following factors play an important role in the success of measurement: the precise control of the lattice sites alignment in the entire three-dimensional space, the uniformity of permanent magnets, and the penetration of magnetic fluxes through waveguide plates.

We first probe the dispersion of antichiral edge states. The numerically calculated band dispersion (obtained using finite-element COMSOL Multiphysics) for a strip with zigzag edges is plotted in Fig. 2(a). It can be found that the projected bulk states (black dots) of the first and second bands form two degenerate points at $k_x = \frac{2\pi}{3a}$ (at 6.9 GHz) and $\frac{4\pi}{3a}$ (at 7.5 GHz). These degenerate points are the projections of K and K′ Dirac points in the bulk. The corresponding Dirac equations near these Dirac points can be obtained through $\mathbf{k}\cdot\mathbf{p}$ analysis (see Supplemental Materials for details), which shows that under the oppositely imposed staggered magnetic flux, the Dirac points only shifts frequencies from K to K′, causing the tilted dispersion of edge states. These edge states are different from the flat edge states in a graphene nanoribbon with zigzag edges [28], because these new edge states propagate both along the +$x$ direction, regardless of the upper or lower edge.

We have constructed a strip sample as shown in Fig. 2(b). Note that the upper (lower) edge corresponds to the A- (B-) type zigzag edge in Fig. 1(c). Copper cladding has been introduced near the upper and lower edges as perfect electric conductor (PEC) boundaries to prevent leakage of electromagnetic waves into the surrounding environment. The left and right edges are covered with microwave absorbers. When a source is located near the upper or lower edge, the excited edge states will always propagate in the +$x$ direction, as experimentally mapped in Figs. 2(c) and (d), respectively. This evidence confirms the existence of antichiral edge states, and is fully consistent with the numerical simulation (see Supplemental Material [26]). After applying Fourier transform to the mapped complex field distribution along the edges (see Supplemental Material [26]), we obtain the edge dispersions at the A- and B-type zigzag edge, as indicated in Figs. 2(e) and (f), respectively, which match well with the computed band dispersion in Fig. 2(a).

These antichiral edge states are topologically protected (see Supplemental Material for the discussion on topological protection), and thus can propagate around certain sharp corners without backscattering. Therefore, it is possible to construct carefully designed finite samples where the antichiral edge states can propagate either fully clockwise or fully counterclockwise along all edges, exhibiting chirality similar to that of conventional chiral edge states.

We consider a sample with a downward triangular shape, as illustrated in Fig. 3(a), whose edges are all A-type zigzag edges. According to the above reasoning, the edge states supported can only propagate clockwise along all edges, exhibiting chirality similar to the conventional chiral edge states. Figure 3(b) shows the photo of a fabricated sample. Since all corners are identical, it is sufficient to demonstrate the robust propagation around one corner. In the measurement, the upper and lower right edges of the sample are confined with copper cladding as PEC boundaries, while the lower left edge is covered with absorbers. A source antenna (labelled as "Port 1" in Fig. 3(b)) is placed near the upper edge while a probing antenna (labelled as "Port 2" in Fig. 3(b)) collects signal after the sharp corner. As shown in Fig. 3(c), the difference between forward and backward transmission is larger than 30 dB around 7.3 GHz, indicating the unidirectionality of edge states. Note that this nonreciprocal-transmission frequency window, 6.9 GHz ~ 7.5 GHz, is consistent with numerical simulation in Fig. 2(a). Then, the field distribution at 7.3 GHz inside the sample has been directly mapped point-by-point, as shown in Fig. 3(d). It can be seen that the edge states are strongly confined to the edges, and can propagate robustly across the 60° sharp corner, with no backscattering observed.

Another sample with an upward triangular shape is illustrated in Fig. 3(e). This sample possess three B-type zigzag edges, and thus shall support edge states propagating counterclockwise along all edges, showing reversed chirality to the situation in Fig. 3(a). Similarly, we fabricate this sample as in Fig. 3(f). The transmission measurement in Fig. 3(g) and field mapping in Fig. 3(h) show similar results to those for the downward-triangular sample, except the reversed direction of propagation for edge states.

In the original proposal in Ref. [20], antichiral edge states emerge at the two parallel edges of a strip sample that extends to infinity without any termination. As a result, the existence of counter-propagating bulk states that balance the directional antichiral edge states is more for the consideration of energy conservation. There is no physical event that can convert antichiral edge states into counter-propagating bulk states. On the other hand, a realistic strip sample shall have certain terminations. It will be interesting to explore the role of such terminations in the energy conversion.

For demonstration purposes, we fabricate a strip sample with a trapezoid termination, whose schematic illustration and real photo are shown in Figs. 4(a) and (b). The upper right 60° corner has no effect on the propagation of antichiral edge states, as we have demonstrated in Fig. 3. That is because the upper and right edges are both A-type zigzag edges. However, the lower right 120° corner can scatter the edge states into backward bulk states. That is because the lower edge is a B-type zigzag edge, which does not match with the other A-type zigzag edges. In the experiment, we place a source near the upper edge to excite the edge states. Field mapping in Fig. 4(c) shows that the edge states can bypass upper right 60° corner, but are fully scattered into bulk states in the backward direction at the lower right 120° corner, without coupling to the lower edge. Note that there is no reflection of edge states observed in the process. Therefore, this energy conversion from the edge states to the bulk states is 100% at the lower right 120° corner.

We have thus directly observed the photonic antichiral edge states in a gyromagnetic photonic crystal. The identical propagation direction along two opposite edges are demonstrated by mapping the dispersion as well as the field distribution of photonic antichiral edge states. Another two unique properties, including the chiral-like robust propagation in samples with certain shapes, and the 100% scattering into backward bulk states at certain terminations, have been further demonstrated in experiment. Although demonstrated in photonics, these results can be extended to other systems for both fermions and bosons, exploring exotic physical phenomena, like Klein tunneling [29], Möbius bands [30], and Andreev reflection [21].

In the final stage of submission, we find a recent preprint that reports the observation of antichiral edge states in a circuit network [31].

This work is supported by National Key Research and Development Program of China (Grant No.2016YFB1200100). Work at Nanyang Technological University is sponsored by Singapore MOE Academic Research Fund Tier 3 Grant MOE2016-T3-1-006, Tier 1 Grants RG187/18 and RG174/16(S), and Tier 2 Grant MOE 2018-T2-1-022(S).

#guigeng001@e.ntu.edu.sg
*denglj@uestc.edu.cn
‡blzhang@ntu.edu.sg

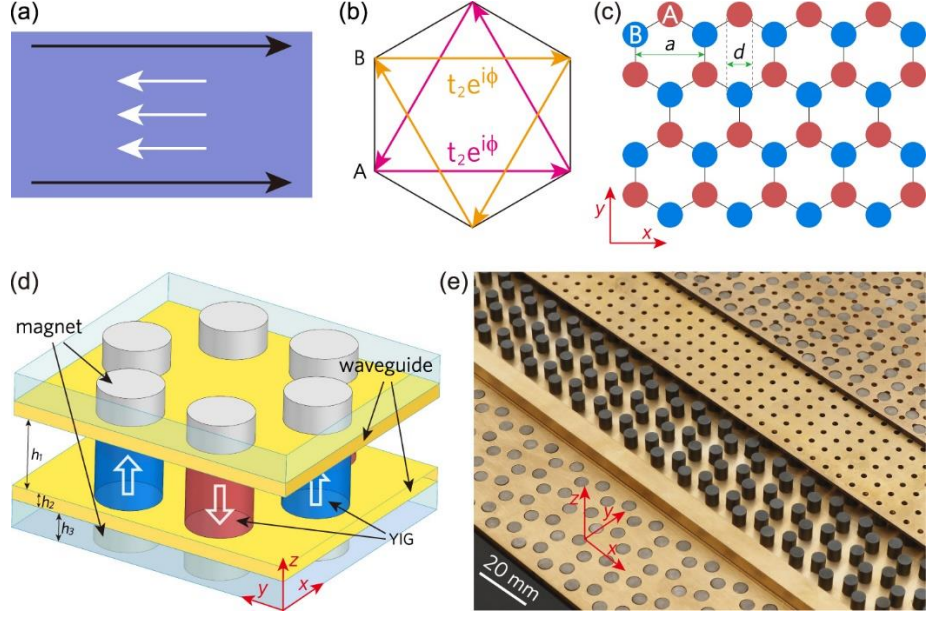

FIG. 1. Buildup of antichiral edge states in a gyromagnetic photonic crystal. (a) Conceptual illustration of antichiral edge states that propagate in the same direction, while the bulk states propagate in the opposite direction. (b) The modified Haldane model. The next-nearest-neighbor coupling $t_2 e^{i\varphi}$ directs inversely in sublattices A and B. (c) A photonic honeycomb lattice consisting of gyromagnetic rods with identical diameter $d = 4.4$ mm. The lattice constant is $a = 12$ mm. The sublattice A (B) indicated in red (blue) is magnetically biased along $+$ $(-)$ $z$ direction. (d-e) Design (d) and implementation (e) of the gyromagnetic photonic crystal. The crystal is placed in a parallel-plate copper waveguide with height $h_1 = 5$ mm. The thickness of waveguide plates is $h_2 = 1$ mm. Permanent magnets of diameter $d = 4.4$ mm and height $h_3 = 2$ mm are embedded in copper plates of the same height, and vertically align with the yttrium iron garnet (YIG) rods. In (e), the compositional layers are glided three lattices away from their edges for photographing.

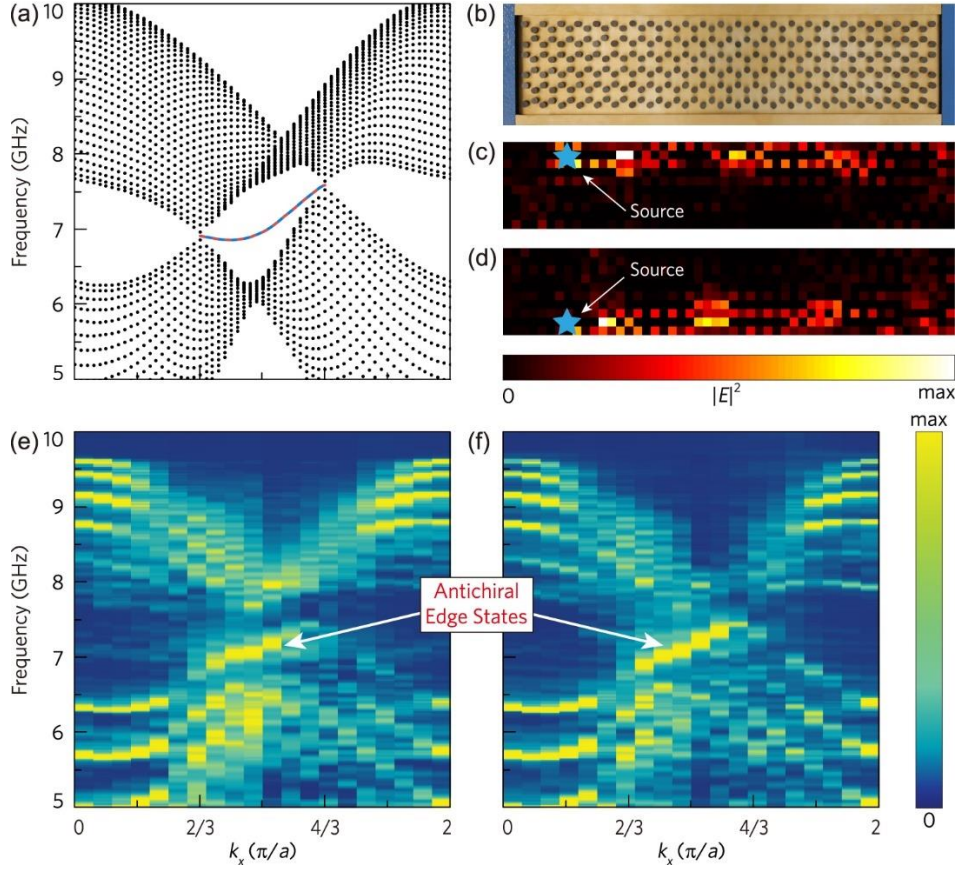

FIG. 2. Observation of antichiral edge states in a strip sample. (a) Numerically calculated bandstructure for a strip with 15 unit cells in *y* direction. The black dots belong to bulk states. (b) Photograph of a strip sample that consists of 5 × 23 unit cells placed on a copper plate. Copper bars are placed at the upper and lower edges as perfect electric conductor (PEC) boundaries. Left and right edges are covered by microwave absorbers (blue). (c-d) Measured field distribution ($|E|^2$) at 7.3 GHz along the upper and lower edges. Blue star shows the position of source antenna. (e-f) Bandstructure obtained from Fourier transform of the edge states mapped in (c) and (d) respectively.

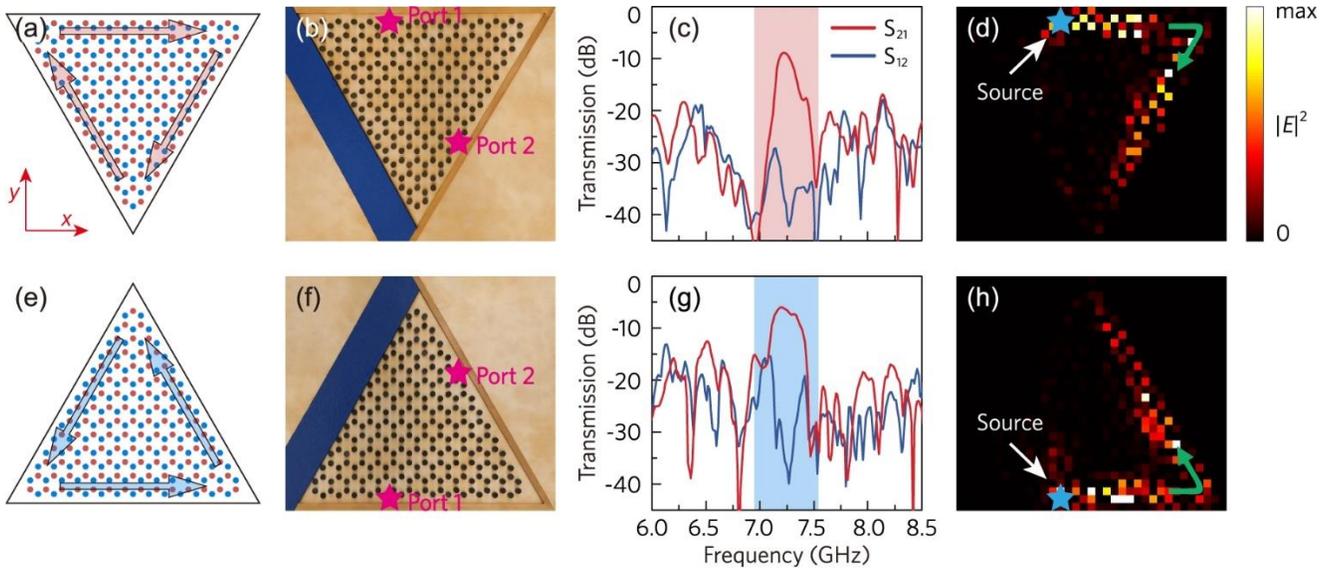

FIG. 3. Robustness of antichiral edge states. (a) Clockwise propagation of antichiral edge states in a downward triangular sample consists of 91 unit cells. All edges are A-type zigzag edges. (b) Photograph of the sample in (a). The sample is wrapped by copper bars and microwave absorber (blue), and placed on a copper plate. (c) Measured transmission. The $S_{12}/S_{21}$ parameter is measured by placing source antenna at port 2/port 1 and detector antenna at port 1/port 2, as indicated by pink stars in (b). (d) Mapped field distribution ($|E|^2$) for the sample at 7.3 GHz. Blue star shows the position of source antenna, as green arrow marks the wave propagation direction. (e) Counterclockwise propagation of antichiral edge states in an upward triangular sample. All edges are B-type zigzag edges. (f) Photograph of the sample in (e). (g-h) Measured results for the sample in (f), with similar labelling to (c) and (d).

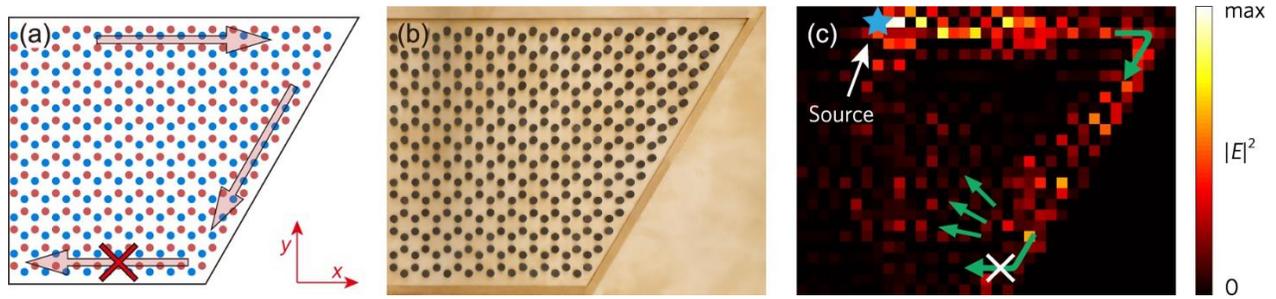

FIG. 4. Conversion of antichiral edge states into bulk states at a termination. (a) Schematic of a strip sample with a trapezoid termination consists of 153 unit cells. (b) Photograph of the sample in (a). The sample is wrapped by copper bars except for the left open edge, and placed on a copper plate. (b) Measured field distributions ($|E|^2$) at 7.3 GHz. Blue star shows the position of source antenna, as green arrows mark the wave propagation direction.
8